\documentclass[aps,draft,amsmath,preprint,floatfix,tightenlines,showpacs]{revtex4}

\usepackage{epsf}

\begin{document}
\title{Anomalous temperature dependence of the supercurrent through a chaotic Josephson junction}

\author{P. W. Brouwer and C. W. J. Beenakker}

\affiliation{Instituut-Lorentz, University of Leiden, 
         P.O. Box 9506, 2300 RA Leiden, 
         The Netherlands}

\begin{abstract}
  We calculate the supercurrent through a Josephson junction consisting of a phase-coherent metal particle (quantum dot), weakly coupled to two superconductors. The classical motion in the quantum dot is assumed to be chaotic on time scales greater than the ergodic time $\tau_{\rm erg}$, which itself is much smaller than the mean dwell time $\tau_{\rm dwell}$. The excitation spectrum of the Josephson junction has a gap $E_{\rm gap}$, which can be less than the gap $\Delta$ in the bulk superconductors. The average supercurrent is computed in the ergodic regime $\tau_{\rm erg} \ll \hbar/\Delta$, using random-matrix theory, and in the non-ergodic regime $\tau_{\rm erg} \gg \hbar/\Delta$, using a semiclassical relation between the supercurrent and dwell-time distribution. 
In contrast to conventional Josephson junctions, raising the temperature above the excitation gap does not necessarily lead to an exponential suppression of the supercurrent. Instead, we find a temperature regime between $E_{\rm gap}$ and $\Delta$ where the supercurrent decreases logarithmically with temperature. This anomalously weak temperature dependence is caused by long-range correlations in the excitation spectrum, which extend over an energy range $\hbar/\tau_{\rm erg}$ greater than $E_{\rm gap} \simeq \hbar/\tau_{\rm dwell}$.
A similar logarithmic temperature dependence of the supercurrent was discovered by Aslamazov, Larkin, and Ovchinnikov, in a Josephson junction consisting of a disordered metal between two tunnel barriers.

\bigskip \noindent
Published in: {\em Chaos, Solitons}\, \& {\em Fractals} {\bf 8}, 1249
(1997). \\
(Pergamon/Elsevier; special issue on Chaos and Quantum Transport in Mesoscopic Cosmos)
  \bigskip
  \pacs{PACS numbers: 74.50.+r, 74.80.Fp, 05.45.+b}
\end{abstract}

\maketitle

\section{Introduction}

The dissipationless flow of a current through a superconductor--normal-metal--superconductor (SNS) junction is a fundamental demonstration of the ``proximity effect'': a normal metal borrows characteristic properties from a nearby superconductor. The energy gap $\Delta$ in the bulk induces a suppression of the density of states inside the normal metal near the Fermi level, depending on the phase difference $\phi$ between the superconductors. The resulting $\phi$-dependence of the free energy $F$ implies the flow of a current $I = (2e/\hbar) d F/d \phi$ in equilibrium. In contrast to the original Josephson effect in tunnel junctions, the separation of the superconductors in an SNS junction can be much greater than the superconducting coherence length. Recent experiments on mesoscopic Josephson junctions \cite{Marsh,Takayanagi,Courtois1995,Courtois1996,Charlat1996,Mur} have revived theoretical interest in this subject\cite{Volkov95,Volkov96,Wilhelm}, which goes back to work by Kulik \cite{Kulik} and Aslamasov, Larkin, and Ovchinnikov \cite{AslamasovLarkinOvchinnikov}. (For more references, see the review \onlinecite{VanWeesTakayanagi}.)

In this paper, we consider the case that the normal region consists of a chaotic quantum dot. A quantum dot is a small metal particle, within which the motion is phase coherent, weakly coupled to the superconductors by means of point contacts. We assume that the classical dynamics in the quantum dot is chaotic on time scales longer than the time $\tau_{\rm erg}$ needed for ergodic exploration of the phase space of the quantum dot. (In order of magnitude, $\tau_{\rm erg} \simeq L/v_{\rm F}$ for a quantum dot of size $L$ without impurities, $v_{\rm F}$ being the Fermi velocity.) On energy scales smaller than $\hbar/\tau_{\rm erg}$, the spectral statistics of a chaotic quantum dot is described by random-matrix theory \cite{Efetov83,AAAS}. On larger energy scales, the non-ergodic dynamics on time scales below $\tau_{\rm erg}$ becomes dominant \cite{AS}. The condition of weak coupling means that the mean dwell time $\tau_{\rm dwell}$ in the quantum dot is much greater than $\tau_{\rm erg}$. (The ratio $\tau_{\rm dwell}/\tau_{\rm erg}$ is of the order of the ratio of the total surface area to the area of the point contacts.) Although Josephson junctions are commonly known as ``weak links'' \cite{Likharev}, we will refer to junctions where $\tau_{\rm dwell} \simeq \tau_{\rm erg}$ as ``strongly coupled'' junctions, to distinguish them from the weakly coupled junctions ($\tau_{\rm dwell} \gg \tau_{\rm erg}$) considered here.

A weakly coupled SNS junction consisting of a dirty normal metal separated from the two superconductors by high tunnel barriers was studied in the original paper by Aslamasov, Larkin, and Ovchinnikov \cite{AslamasovLarkinOvchinnikov}. Their theory was restricted to the high-temperature regime $kT \gg \hbar/\tau_{\rm dwell}$. In contrast to strongly coupled Josephson junctions, where the supercurrent is suppressed exponentially for $kT \gtrsim \hbar/\tau_{\rm dwell}$ \cite{Wilhelm,Svidzinskii}, it was found that the supercurrent depends logarithmically on temperature for $k T \lesssim \hbar/\tau_{\rm erg}$, while exponential suppression only sets in when $k T \gtrsim \hbar/\tau_{\rm erg}$. In the present paper we find a qualitatively similar high-temperature behavior of the supercurrent in the case that the weak coupling is ensured by point contacts rather than tunnel barriers. In addition, we are able to go down to zero temperature, where we find that the supercurrent acquires a logarithmic dependence on the minimum of $\tau_{\rm dwell}/\tau_{\rm erg}$ and $\tau_{\rm dwell} \Delta/\hbar$, over and above the conventional dependence on $\min(\hbar/\tau_{\rm dwell},\Delta)$ known from strongly-coupled Josephson junctions.

Our paper builds on earlier work with Melsen and Frahm \cite{Melsen96}, where we computed the density of states of a chaotic quantum dot which is weakly coupled to a superconductor, and found a gap at the Fermi level of width $E_{\rm gap} \simeq \hbar/\tau_{\rm dwell}$ (provided $\hbar/\tau_{\rm dwell} \ll \Delta$). Although the supercurrent can be expressed as an integral over the density of states, direct application of the results of Ref.\ \onlinecite{Melsen96} is not possible, as they were only derived under the condition $\varepsilon \lesssim \hbar/\tau_{\rm dwell}$. We have it found it necessary to relax this restriction, because the sensitivity of the excitation spectrum to the phase difference of the superconductors (which determines the supercurrent) extends up to $\hbar/\tau_{\rm erg}$. It is because of this long-range sensitivity, that the temperature scale for the exponential suppression of the supercurrent is set by $\tau_{\rm erg}$ and not by $\tau_{\rm dwell}$.

This work is organized as follows: In Sec.\ \ref{sec:2} we review the scattering theory of the Josephson effect \cite{Beenakker91,Beenakker92}, on which our calculation is based. We distinguish two regimes, the ergodic and non-ergodic regime, depending on the relative magnitude of $\Delta$ and $\hbar/\tau_{\rm erg}$. In Secs.\ \ref{sec:3} and \ref{sec:4}, we consider the ergodic regime $\tau_{\rm erg} \ll \hbar/\Delta$, where we can use random-matrix theory. The non-ergodic regime $\tau_{\rm erg} \gtrsim \hbar/\Delta$ is treated in Sec.\ \ref{sec:5}, using a semiclassical relation between the supercurrent and the dwell-time distribution. We conclude with an overview of our results in Sec.\ \ref{sec:6}.

\section{Scattering matrix formula for the supercurrent} \label{sec:2}

\begin{figure}
\hspace{0.3\hsize}
\epsfxsize=0.37\hsize
\epsffile{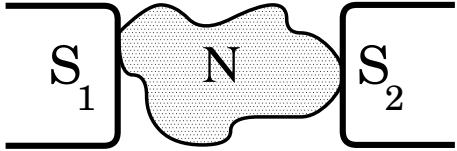}
\caption{\label{fig:1} Schematic drawing of a chaotic Josephson junction.}
\end{figure}

We consider the SNS junction sketched in Fig.\ \ref{fig:1}. The two superconductors S$_1$ and S$_2$ have order parameters $\Delta e^{\phi_1}$, $\Delta e^{i \phi_2}$, with phase difference $\phi = \phi_1 - \phi_2$. The contacts to the normal metal N have $N_1$, $N_2$ propagating modes at the Fermi energy $E_{\rm F}$. We denote $N=N_1+N_2$. Elastic scattering by the normal metal at energy $E = E_{\rm F} + \varepsilon$ is characterized by an $N \times N$ unitary matrix $S(\varepsilon)$. Excitations in N with energy $\varepsilon>0$ consist of electrons (occupied states lying $\varepsilon$ above the Fermi level) and holes (empty states lying $\varepsilon$ below the Fermi level). Their scattering matrix ${\cal S}_{\rm N}(\varepsilon)$ has dimension $2N \times 2N$, with the block structure
\begin{equation}
 {\cal S}_{\rm N}(\varepsilon) =
  \left(\begin{array}{cc} S(\varepsilon) & 0 \\ 0 & S^{*}(-\varepsilon) \end{array} \right) \equiv
  \left(\begin{array}{cc} S^{ee} & S^{eh} \\ S^{he} & S^{hh} \end{array}\right). \label{eq:SN}
\end{equation}
The off-diagonal blocks are zero, because the normal metal does not scatter electrons into holes.

The supercurrent couples electron and hole excitations through the mechanism of Andreev reflection \cite{Andreev}: An electron approaching the NS interface from the normal side at $\varepsilon < \Delta$ is reflected as a hole, and vice versa. The scattering matrix ${\cal S}_{\rm A}(\varepsilon)$ for Andreev reflection is given by (assuming $\Delta \ll E_{\rm F}$)
\begin{subequations} \label{eq:SA}
\begin{eqnarray} 
  {\cal S}_{\rm A}(\varepsilon) &=& \alpha(\varepsilon) 
    \left(\begin{array}{cc} 0 & e^{i \Phi} \\
                            e^{-i \Phi} & 0\end{array}\right), \\
  \alpha(\varepsilon) &=& e^{-i \arccos(\varepsilon/\Delta)}
  = \varepsilon/\Delta - i \sqrt{1 - \varepsilon^2/\Delta^2}.\end{eqnarray}
\end{subequations}%
Here $\Phi$ is a diagonal matrix with elements $\Phi_{jj} = \phi_1 = \phi/2$ for $1 \le j \le N_1$ and $\Phi_{jj} = \phi_2 = -\phi/2$ for $N_1+1 \le j \le N_1+N_2$. The matrix ${\cal S}_{\rm A}$ has been defined for $\varepsilon < \Delta$. Its definition may be extended to $\varepsilon > \Delta$, when $\alpha(\varepsilon) = \varepsilon/\Delta - \sqrt{\varepsilon^2-\Delta^2}$. Notice that ${\cal S}_{\rm A}$ is no longer unitary for $\varepsilon > \Delta$.

The matrices ${\cal S}_{\rm N}$ and ${\cal S}_{\rm A}$ determine the excitation spectrum of the Josephson junction for all $\varepsilon$, in the following way \cite{Beenakker91}. For $\varepsilon < \Delta$, the spectrum is discrete. The density of states $\rho(\varepsilon) = \sum_{n} \delta(\varepsilon - \varepsilon_{n})$ consists of delta functions at the solutions of the equation 
\begin{equation}
  \det\left[1 - {\cal S}_{\rm A}(\varepsilon_n) {\cal S}_{\rm N}(\varepsilon_n)\right] = 0.
\end{equation}
For $\varepsilon > \Delta$, the spectrum is continuous, with density of states
\begin{equation}
  \rho(\varepsilon) = -{1 \over \pi} \mbox{Im}\,
    {d \over d \varepsilon} \left[
    \ln \det \left[1-{\cal S}_{\rm A}(\varepsilon) {\cal S}_{\rm N}(\varepsilon)\right]
    - {1 \over 2} \ln \det {\cal S}_{\rm A} (\varepsilon) {\cal S}_{\rm N}(\varepsilon) \right].
  \label{eq:RhoSEq}
\end{equation}
If $\varepsilon$ is replaced by $\varepsilon + i 0^{+}$, Eq.\ (\ref{eq:RhoSEq}) describes both the continuous and the discrete spectrum \cite{DoronSmilansky}. The excitation spectrum determines the free energy $F$ of the Josephson junction,
\begin{equation}
  F = - 2 k T \int_0^{\infty} d\varepsilon\, \rho(\varepsilon) 
    \ln \left[ 2 \cosh(\varepsilon/2 k T) \right]
    \ + \ [\mbox{$\phi$-independent terms}], \label{eq:F}
\end{equation}
where $T$ is the temperature. The $\phi$-independent terms include \cite{BeenakkerVanHouten} a spatial integral over $|\Delta(\vec r)|^2$, which does not depend on $\phi$ in the step-function model for the pair potential $\Delta(\vec r)$. Only the $\phi$-dependent terms contribute to the supercurrent
\begin{equation}
  I = {2 e \over \hbar} {dF \over d\phi}. \label{eq:IF}
\end{equation}

We use the analyticity of ${\cal S}_{\rm A}$ and ${\cal S}_{\rm N}$ in the upper half of the complex $\varepsilon$-plane to rewrite the expression for the supercurrent in a more convenient form. Under a change $\varepsilon \to - \varepsilon$, the determinant $\det[1-{\cal S}_{\rm A}(\varepsilon+i0^{+}) {\cal S}_{\rm N}(\varepsilon+i0^{+})]$ goes over into its complex conjugate. Combination of Eqs.\ (\ref{eq:RhoSEq})--(\ref{eq:IF}), and extension of the $\varepsilon$-integration from $-\infty$ to $\infty$, results in
\begin{equation} \label{eq:IF2}
  I = {e \over i \pi \hbar} 2 k T {d \over d \phi} \int_{-\infty+i0^{+}}^{\infty+i0^{+}}
   d \varepsilon
   \ln\left[ 2 \cosh(\varepsilon/2 k T) \right] 
   {d \over d \varepsilon}
   \ln \det[1-{\cal S}_{\rm A}(\varepsilon) {\cal S}_{\rm N}(\varepsilon)].
\end{equation}
We now perform a partial integration and close the integration contour in the upper half of the complex plane. The integrand has poles at the Matsubara frequencies $i \omega_n = (2n+1) i \pi k T $. Summing over the residues one finds
\begin{equation}
  I = -{2e \over \hbar} 2 k T {d \over d \phi}
    \sum_{n=0}^{\infty} \ln \det[1-{\cal S}_{\rm A}(i \omega_n)
      {\cal S}_{\rm N}(i \omega_n)]. \label{eq:Iomegaformula}
\end{equation}
Eq.\ (\ref{eq:Iomegaformula}) is the starting point for our evaluation of the average supercurrent through a chaotic Josephson junction.

\section{Supercurrent through a chaotic Josephson junction} \label{sec:3}

We consider the case that the normal region has a chaotic classical dynamics on time scales greater than the ergodic time $\tau_{\rm erg}$. In this section, we assume that $\tau_{\rm erg} \ll \hbar/\Delta$, so that we may use random-matrix theory to evaluate the ensemble average of the supercurrent. We postpone to Sec.\ \ref{sec:5} a discussion of the regime $\tau_{\rm erg} \gtrsim \hbar/\Delta$, in which the non-ergodic dynamics on time scales shorter than $\tau_{\rm erg}$ starts to play a role. We assume that the normal metal is weakly coupled to the superconductors, so that the mean dwell time $\tau_{\rm dwell} \gg \tau_{\rm erg}$. No assumption is made regarding the relative magnitudes of $\tau_{\rm dwell}$ and $\hbar/\Delta$. 

We use a relationship between the scattering matrix $S$ of the normal metal and its Hamiltonian $H$ \cite{MahauxWeidenmueller,VWZ},
\begin{equation} \label{eq:Sformula}
  S(\varepsilon) = 1 - 2\pi i W^{\dagger}(\varepsilon-H+i \pi W W^{\dagger})^{-1}W.
\end{equation}
The Hamiltonian $H$ (representing the isolated normal metal region) is taken from the Gaussian ensemble of random-matrix theory \cite{Mehta},
\begin{equation}
  P(H) \propto \exp\left(-\frac{1}{4} M \lambda^{-2} \mbox{tr}\, H^2\right),
\end{equation}
where $M$ is the dimension of $H$ (taken to infinity at the end) and $\lambda$ is a parameter that determines the average level spacing $\delta=\lambda \pi/2M$ of the excitation spectrum in the normal region. (This spacing $\delta$ is half the level spacing of $H$, because it combines electron and hole levels together.) The matrix $H$ is real and symmetric. The coupling matrix $W$ is an $M \times N$ matrix with elements\cite{IWZ,Brouwer95}
\begin{eqnarray}
  W_{mn} &=& {1 \over \pi} \delta_{mn} (2 M \delta)^{1/2} \left(2 \Gamma_n^{-1} - 1 
  - 2 \Gamma_n^{-1} \sqrt{1 - \Gamma_n} \right)^{1/2}. \label{eq:W}
\end{eqnarray}
Here $\Gamma_n$ is the transmission probability of mode $n$ in the contacts to the superconductor. For ballistic contacts, $\Gamma_n = 1$, while $\Gamma_n \ll 1$ for tunneling contacts. 

We now substitute Eq.\ (\ref{eq:Sformula}) for $S$ into Eq.\ (\ref{eq:SN}) for ${\cal S}_{\rm N}$ and then substitute ${\cal S}_{\rm N}$ into Eq.\ (\ref{eq:Iomegaformula}) for the supercurrent. Using also Eq.\ (\ref{eq:SA}) for ${\cal S}_{\rm A}$, we find after some straightforward matrix algebra that
\begin{eqnarray}
  I &=&
  -{2e \over \hbar} 2 k T {d \over d \phi} \sum_{n=0}^{\infty}
    \ln \det[i \omega_n - {\cal H} + {\cal W}(i \omega_n)],
\end{eqnarray}
where we have introduced the $2M \times 2M$ matrices 
\begin{subequations}
\begin{eqnarray}
  {\cal H} &=&
    \left(\begin{array}{cc} H & 0 \\ 0 & -H \end{array}\right),\\ 
  {\cal W}(\varepsilon) &=& 
    {\pi \Delta \over \sqrt{\Delta^2 - \varepsilon^2}} 
      \left(\begin{array}{cc} 
      (\varepsilon/\Delta)  W W^{\rm T} &
      W e^{i\Phi^{\vphantom{M}}} W^{\rm T}\\
      W e^{-i\Phi} W^{\rm T} &
      (\varepsilon/\Delta) W W^{\rm T} \end{array}\right). \label{eq:Hdef}
\end{eqnarray}
\end{subequations}%
The matrix ${\cal H} - {\cal W}(\varepsilon)$ is the effective Hamiltonian of Refs.\ \onlinecite{Melsen96,Frahm96} (where the regime $\varepsilon \ll \Delta$ was considered, in which the $\varepsilon$-dependence of ${\cal W}(\varepsilon)$ can be neglected).

We define the $2M \times 2M$ Green function
\begin{eqnarray}
  {\cal G}(\varepsilon) &=& \left[\varepsilon - {\cal H} + {\cal W}(\varepsilon)\right]^{-1},
\end{eqnarray}
which determines the density of states according to
\begin{eqnarray}
  \rho(\varepsilon) &=& -\pi^{-1} \mbox{Im}\, \mbox{tr}\,  {\cal
    G}(\varepsilon+i0)
  \left( 1 + \frac{\partial {\cal W}}{\partial \varepsilon} \right).
  \label{eq:RhoGreen}
\end{eqnarray}
Eq.\ (\ref{eq:RhoGreen}) is equivalent to Eq.\ (\ref{eq:RhoSEq}). The expression for the supercurrent in terms of ${\cal G}(\varepsilon)$ is
\begin{eqnarray}
  I &=& {2e \over \hbar} 2 k T {d \over d \phi} \sum_{n=0}^{\infty}
    \ln \det {\cal G}(i \omega_n) \nonumber  \\ &=&
    -{2e \over \hbar} 2 k T \sum_{n=0}^{\infty} \mbox{tr}\, 
    {\cal G}(i \omega_n)
    {d \over d \phi}  {\cal W}(i \omega_n).
\end{eqnarray}

The average supercurrent follows from the average Green function $\langle {\cal G}(\varepsilon) \rangle$, since ${\cal W}$ is a fixed matrix. The average over the random Hamiltonian $H$ (determining ${\cal G}$) is done with the help of the diagrammatic technique of Refs.\ \onlinecite{Pandey,BrezinZee}. We consider the regime $M, N, |\varepsilon|/\delta \gg 1$ in which only planar diagrams need to be considered. Resummation of these diagrams leads to a self-consistency equation which is similar to Pastur's equation \cite{Pastur},
\begin{equation} \label{eq:Gdyson}
  \langle {\cal G}(\varepsilon) \rangle 
  = \left[ \vphantom{M_M^M} \varepsilon + {\cal W}(\varepsilon)
    - (\lambda^2/M) {\cal P}(\varepsilon) \otimes {\openone_{M}} \right]^{-1}.
\end{equation}
The symbol $\otimes$ indicates the direct product between the $M \times M$ unit matrix $\openone_{M}$ and the $2 \times 2$ matrix
\begin{equation} \label{eq:PDef}
  {\cal P} = \left( \begin{array}{cc} 
    \langle \mbox{tr}\, {\cal G}^{ee} \rangle & - \langle \mbox{tr}\, {\cal G}^{eh} \rangle \\
    - \langle \mbox{tr}\, {\cal G}^{he} \rangle & \langle \mbox{tr}\, {\cal G}^{hh} \rangle
    \end{array} \right).
\end{equation}
We seek the solution of Eqs.\ (\ref{eq:Gdyson})--(\ref{eq:PDef}) which satisfies 
\begin{equation} \label{eq:PastBound}
  \varepsilon {\cal G}(\varepsilon) \to \openone_{2M}\ \mbox{if $|\varepsilon| \gg \lambda$}.
\end{equation}

It is convenient to define a self energy
\begin{equation} \label{eq:SigmaDef}
  {\bf \Sigma} = \left( \begin{array}{cc} 
    {\bf \Sigma}^{ee} &  {\bf \Sigma}^{eh} \\
    {\bf \Sigma}^{he} &  {\bf \Sigma}^{hh} \end{array} \right) =
    {\lambda \over M} \left( \begin{array}{cc} 
    \mbox{tr}\, {\cal G}^{ee} &  \mbox{tr}\, {\cal G}^{eh} \\
    \mbox{tr}\, {\cal G}^{he} &  \mbox{tr}\, {\cal G}^{hh} \end{array} \right).
\end{equation}
Eqs.\ (\ref{eq:Gdyson})--(\ref{eq:SigmaDef}) contain a closed set of equations from which $\langle {\bf \Sigma} \rangle$ can be determined. We are interested in the limit $M \to \infty$, $\lambda \to \infty$, keeping $N$ and $\varepsilon/\delta = 2 \varepsilon M/\lambda\pi$ fixed. In this limit, the equations for $\langle {\bf \Sigma} \rangle$ become
\begin{subequations} \label{eq:SigmaEqSet}
\begin{eqnarray}\label{eq:SigmaEqSeta}
  && \langle {\bf \Sigma}^{ee} \rangle = \langle {\bf \Sigma}^{hh} \rangle ,\ \ \
  \langle {\bf  \Sigma}^{eh} \rangle
    \langle {\bf  \Sigma}^{he} \rangle -
    \langle {\bf  \Sigma}^{ee} \rangle^2 = 1, \\ \label{eq:SigmaEqSetb}
  && \frac{\pi \varepsilon}{2\delta}
    \langle {\bf  \Sigma}^{eh} \rangle +
  \sum_{j=1}^{N} K_j \left( {\varepsilon \langle {\bf  \Sigma}^{eh} \rangle +
    \Delta e^{ i \Phi_{jj}} \langle {\bf  \Sigma}^{ee} \rangle} \right)
    = 0, \\ \label{eq:SigmaEqSetc}
  && \frac{\pi \varepsilon}{2\delta}
    \langle {\bf  \Sigma}^{he} \rangle + 
  \sum_{j=1}^{N} K_j \left( {\varepsilon \langle {\bf  \Sigma}^{he} \rangle +
    \Delta  e^{-i \Phi_{jj}} \langle {\bf  \Sigma}^{ee} \rangle} \right)
    = 0.
\end{eqnarray}
The function $K_j(\varepsilon)$ is defined through
\begin{eqnarray} \label{eq:SigmaEqSetd}
  && \Gamma_{j}/K_j = (4 - 2\Gamma_{j}) \sqrt{\Delta^2 - \varepsilon^2} + \Gamma_{j}
     \left(
     \Delta e^{-i \Phi_{jj}}
       \langle {\bf  \Sigma}^{eh} \rangle +
     \Delta e^{ i \Phi_{jj}}
       \langle {\bf  \Sigma}^{he} \rangle +
     2 \varepsilon \langle {\bf  \Sigma}^{ee} \rangle \right).
\end{eqnarray}
\end{subequations}%
(We substituted Eq.\ (\ref{eq:W}) for the matrix $W$.) The boundary condition (\ref{eq:PastBound}) becomes ineffective in the limit $\lambda \to \infty$. Instead, we seek the solution of Eq.\ (\ref{eq:SigmaEqSet}) with $\langle {\bf \Sigma}^{ee} \rangle = \langle {\bf \Sigma}^{hh} \rangle \to -i$ for $\varepsilon \to i \infty$, corresponding to a constant density of states $\rho(\varepsilon) = 1/\delta$ for $|\varepsilon| \gg \Delta$. From $\langle {\bf \Sigma} \rangle$, we find $\langle {\cal G} \rangle$ and hence the ensemble averaged supercurrent $\langle I \rangle$,
\begin{eqnarray} \label{eq:SigmaSuper}
  \langle I \rangle &=& {2 e \over i \hbar} k T \Delta \sum_{n=0}^{\infty} \sum_{j=1}^{N} \mbox{sign}(\Phi_{jj})\, K_j \left[ e^{i \Phi_{jj}} \langle {\bf \Sigma}^{he}(i \omega_n) \rangle - e^{-i \Phi_{jj}} \langle {\bf \Sigma}^{eh}(i \omega_n) \rangle \right].
\end{eqnarray}
Eqs.\ (\ref{eq:SigmaEqSet}) and (\ref{eq:SigmaSuper}) contain all the information needed to determine the average supercurrent through a chaotic Josephson junction.

An analytic solution of Eq.\ (\ref{eq:SigmaEqSet}) is possible in certain limiting cases. Here we discuss the case of high tunnel barriers, $\Gamma_{j} \ll 1$ for all $j$. Then we may approximate $K_j = (1/4) \Gamma_j (\Delta^2-\varepsilon^2)^{-1/2}$ and find
\begin{subequations} \label{eq:TunnelSigma}
\begin{eqnarray}
  \langle {\bf  \Sigma}^{ee} \rangle &=& \langle {\bf  \Sigma}^{hh} \rangle =
  {-\varepsilon (\sqrt{\Delta^2 - \varepsilon^2} + E_{\rm T}) \left[{|\Omega|^2 \Delta^2 - \varepsilon^2 \left(\sqrt{\Delta^2 - \varepsilon^2} + E_{\rm T}\right)^2}\right]^{-{1 \over 2}}}, \\
  \langle {\bf  \Sigma}^{eh} \rangle &=& 
  {\Omega\hphantom{^{*}} \Delta \left[{|\Omega|^2 \Delta^2 - \varepsilon^2 \left(\sqrt{\Delta^2 - \varepsilon^2} + E_{\rm T}\right)^2}\right]^{-{1 \over 2}}},\\
  \langle {\bf  \Sigma}^{he} \rangle &=& 
  {\Omega^{*} \Delta \left[{|\Omega|^2 \Delta^2 - \varepsilon^2 \left(\sqrt{\Delta^2 - \varepsilon^2} + E_{\rm T}\right)^2}\right]^{-{1 \over 2}}}, \\
  \Omega(\phi) &=& {\delta \over 2 \pi} \sum_{j=1}^{N} \Gamma_j e^{i \Phi_{jj}}, \ \ E_{\rm T} = {\delta \over 2 \pi} \sum_{j=1}^{N} \Gamma_j = \Omega(0). \label{eq:Ed}
\end{eqnarray}
\end{subequations}%
The energy $E_{\rm T}$ is related to the mean dwell time through $E_{\rm T} = \hbar/2 \tau_{\rm dwell}$ \cite{Dwell}. The excitation gap in the spectrum of the Josephson junction is of order $|\Omega(\phi)|$ when $\tau_{\rm dwell} \gg \hbar/\Delta$ \cite{Melsen96}. Substitution of Eq.\ (\ref{eq:TunnelSigma}) into Eq.\ (\ref{eq:SigmaSuper}) yields the supercurrent
\begin{eqnarray} \label{eq:Itunnel}
  \langle I \rangle &=& {2 \pi \over e} k T G E_{\rm T} \sum_{n=0}^{\infty} {\Delta^2 \sin\phi \over \sqrt{\Delta^2 + \omega_n^2}} \left[{|\Omega|^2 \Delta^2 + \omega_n^2 \left(\sqrt{\Delta^2 + \omega_n^2} + E_{\rm T}\right)^2}\right]^{-{1 \over 2}},
\end{eqnarray}
where 
\begin{equation}
  G = {2e^2 \over h}{\sum_{i=1}^{N_1} \sum_{j=N_1+1}^{N} \Gamma_i\Gamma_j \over \sum_{k=1}^{N} \Gamma_k} \label{eq:G}
\end{equation}
is the conductance of the Josephson junction when the superconductors are in the normal state.

\begin{figure}

\hspace{0.2\hsize}
\epsfxsize=0.55\hsize
\epsffile{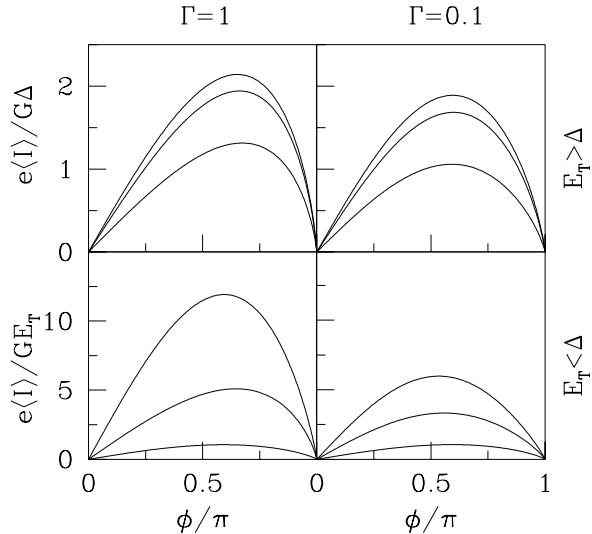}
\par

\caption{\label{fig:3} Average supercurrent at zero temperature, computed from Eqs.\ (\protect\ref{eq:SigmaEqSet}) and (\protect\ref{eq:SigmaSuper}) for the case $N_1 = N_2 = N/2$, $\Gamma_j = \Gamma$ for all $j$. Left panels: $\Gamma=1$; right panels: $\Gamma=0.1$. The upper panels show $\langle I \rangle$ in the short dwell-time regime for $E_{\rm T}/\Delta= 1$ (bottom curve), $10$, and $100$ (top curve). The bottom panels show $\langle I \rangle$ in the long dwell-time regime for $E_{\rm T}/\Delta = 0.01$ (top curve), $0.1$, and $1$ (bottom curve). The conductance $G = (2e^2/h) N \Gamma/4$ and the Thouless energy $E_{\rm T} = N\Gamma\delta/2\pi$. Notice that $\langle I \rangle$ is in units of $G \Delta/e$ in the top panels, and in units of $G E_{\rm T}/e$ in the bottom panels.}
\end{figure}

For arbitrary transmission probabilities $\Gamma_j$, it is necessary to solve Eq.\ (\ref{eq:SigmaEqSet}) numerically. We have studied the case that both point contacts have an equal number of modes ($N_1 = N_2 = N/2$), and that all transmission probabilities are equal ($\Gamma_j = \Gamma$ for all $j$). The average supercurrent at zero temperature for $\Gamma=0.1$ and $\Gamma=1$ is shown in Fig.\ \ref{fig:3}.

\section{Ergodic regime} \label{sec:4}

The general result (\ref{eq:SigmaEqSet})--(\ref{eq:SigmaSuper}) describes the supercurrent in the ergodic regime $\tau_{\rm erg} \ll \hbar/\Delta$. Within this regime, we can distinguish two further regimes, depending on whether the dwell time $\tau_{\rm dwell} = \hbar/2E_{\rm T}$ is short or long compared to $\hbar/\Delta$. We discuss these two regimes in two separate subsections.

\subsection{Short dwell-time regime}

In the short dwell-time regime (when $\tau_{\rm dwell} \ll \hbar/\Delta$, or equivalently $E_{\rm T} \gg \Delta$), the magnitude of the critical current $I_{\rm c} = \max_{\phi} I(\phi)$ is set by the energy gap $\Delta$ in the bulk superconductor: $I_{\rm c} \simeq G \Delta/e$ at zero temperature. The temperature dependence of $I_{\rm c}$ can be neglected as long as $k T\ll \Delta$, i.e.\ for temperatures $T$ much less than the critical temperature $T_{\rm c}$ of the bulk superconductor. In the case of tunneling contacts, evaluation of
Eq.\ (\ref{eq:Itunnel}) with $E_{\rm T} \gg \Delta \gg k T$ yields
\begin{eqnarray}
  \langle I \rangle = {G \Delta \over e} {K\biglb(\gamma \sin(\phi/2)\bigrb) \sin \phi \over \sqrt{1 - \gamma^2 \sin^2(\phi/2)}}. \label{eq:Short}
\end{eqnarray}
The conductance $G$ was defined in Eq.\ (\ref{eq:G}), the function $K$ is the complete elliptic integral of the first kind, and we abbreviated 
\begin{equation} \label{eq:y}
  \gamma = 2 \left( \sum_{i=1}^{N_1} \sum_{j=N_1+1}^{N} \Gamma_i \Gamma_j \right)^{1/2} \left( \sum_{k=1}^{N} \Gamma_k \right)^{-1}.
\end{equation}
The parameter $\gamma$ equals $1$ for two identical point contacts with mode-independent tunnel probabilities.

The result (\ref{eq:Short}) could also have been obtained directly from the general formula for the zero-temperature supercurrent in the short dwell-time regime \cite{Beenakker91},
\begin{equation}
  \langle I \rangle = {e \Delta \over 2 \hbar} \int_0^1 dt\, \rho(t) {t \sin \phi \over \sqrt{1 - t \sin^2(\phi/2)}}, \label{eq:Irho}
\end{equation}
which relates $\langle I \rangle$ to an integral over the transmission eigenvalues $t$ of the junction in the normal state, with density $\rho(t)$. The transmission eigenvalue density for a chaotic cavity with two identical tunneling contacts ($N_1 = N_2 = N/2$, $\Gamma_j = \Gamma_{j+N/2}$ for $j=1,2,\ldots,N/2)$ is given by \cite{BrouwerBeenakker1996a}
\begin{equation}
  \rho(t) = \sum_{j=1}^{N/2} {\Gamma_j(2-\Gamma_j) \over \pi (\Gamma_j^2 - 4 \Gamma_j t + 4 t) \sqrt{t(1-t)}}.
\end{equation}
One can check that the integral (\ref{eq:Irho}) equals Eq.\ (\ref{eq:Short}) with $\gamma=1$ if $\Gamma_j \ll 1$ for all $j$. For two identical ballistic point contacts ($N_1=N_2=N/2$, $\Gamma_j = 1$ for all $j$), the density is $\rho(t) = N (2 \pi)^{-1} [t(1-t)]^{-1/2}$ \cite{BarangerMello94,JPB}, which yields
\begin{equation}
  \langle I \rangle = 
 {2 e \Delta \over i \pi \hbar \sin(\phi/2)} G E\Biglb(i\, \mbox{arsinh}[ \tan(\phi/2)],i\, \mbox{cotan}(\phi/2)\Bigrb).
\end{equation}
Here $G=N/4$ and $E$ is the elliptic integral of the second kind.

\subsection{Long dwell-time regime}

In the long dwell-time regime (when $\tau_{\rm dwell} \gg \hbar/\Delta$, or equivalently $E_{\rm T} \ll \Delta$), the magnitude of the critical current is set by the Thouless energy, but retains a logarithmic dependence on $\Delta$: $I_{\rm c} \simeq (G E_{\rm T}/e) \ln(\Delta/E_{\rm T})$. The temperature dependence of $I_{\rm c}$ can be neglected as long as $kT \ll E_{\rm T}$. If $k T \gg E_{\rm T}$ (but still $T \ll T_{\rm c}$) the critical current decreases, though only logarithmically: $I_{\rm c} \simeq (G E_{\rm T}/e) \ln(\Delta/k T)$. For the case of tunneling contacts, we find from Eq.\ (\ref{eq:Itunnel}) the expressions
\begin{subequations}\label{eq:LongRes}
\begin{eqnarray}
  \langle I \rangle &=& {G E_{\rm T} \over e} \sin \phi 
    {\, \ln \left( {2\Delta / E_{\rm T} \over \sqrt{1 - \gamma^2 \sin^2(\phi/2)}} \right) } \ \  k T \ll E_{\rm T}, \\
  \langle I \rangle &=& {G E_{\rm T} \over e} \sin \phi
    \lefteqn{\, \left[ \vphantom{M^M_M} \ln \left( {2\Delta \over \pi k T} \right) + c_{\rm Euler} \right]}
    \hphantom{\, \ln \left( { 2\Delta / E_{\rm T} \over \sqrt{1 - \gamma^2 \sin^2(\phi/2)}} \right) } \ \
    k T \gg E_{\rm T},\ T \ll T_{\rm c},
\end{eqnarray}
\end{subequations}%
where $c_{\rm Euler} \approx 0.58$ is Euler's constant. For ballistic contacts, we do not have a simple expression as Eq.\ (\ref{eq:LongRes}), but the parametric dependence of $I$ on $\Delta$, $E_{\rm T}$, and $k T$ is the same as for tunneling contacts (cf.\ Fig.\ \ref{fig:3}).

The logarithmic dependence on $\Delta$ of the supercurrent when $E_{\rm T} \ll \Delta$ arises because the Thouless energy $E_{\rm T}$ is not an effective cutoff for the Matsubara sum (\ref{eq:Iomegaformula}) or, equivalently, for the energy integration (\ref{eq:IF2}). Spectral correlations exist up to energies of order $\hbar/\tau_{\rm erg} \gg E_{\rm T}$. These long-range spectral correlations are responsible for the weak decay ${\bf \Sigma}^{eh} \propto 1/\omega$ of the self-energy and $\rho-\delta^{-1} \propto 1/\varepsilon^2$ of the density of states. The superconducting energy gap $\Delta$ has to serve as a cutoff energy for the otherwise logarithmically divergent Eqs.\ (\ref{eq:IF2}) and (\ref{eq:Iomegaformula}), which explains the logarithm $\ln \Delta$ in Eq.\ (\ref{eq:LongRes}).

\section{Non-ergodic regime} \label{sec:5}

When $\tau_{\rm erg} \gtrsim \hbar/\Delta$, a random-matrix theory of the Josephson effect is no longer possible, because the non-ergodic dynamics on time scales shorter than $\tau_{\rm erg}$ starts to play a role. To study the average supercurrent in this non-ergodic regime, we return to Eq.\ (\ref{eq:Iomegaformula}). On substitution of Eqs.\ (\ref{eq:SN}) and (\ref{eq:SA}) we obtain an expression for $I$ in terms of the scattering matrix $S$ of the normal region,
\begin{subequations} \label{eq:IFsum}
\begin{eqnarray}
  I &=& {2 \pi \over e} k T \sum_{n=0}^{\infty} F(\omega_n), 
    \\ F(\omega) &=& - {4e^2 \over h} {d \over d \phi} \mbox{tr}\, 
   \ln [1-\alpha(i \omega)^2 S(i \omega) e^{i \Phi} S^{*}(-i \omega) e^{-i \Phi}]. \label{eq:Iomegaformula2}
\end{eqnarray}
\end{subequations}%

The evaluation of the scattering matrix at the imaginary energy $i \omega_n$ is equivalent to the evaluation of the scattering matrix at the Fermi level in the presence of absorption, with rate $1/\tau_{\rm abs} = 2 \omega_n/\hbar = (2n+1) 2 \pi k T/\hbar$. We first consider temperatures $k T \gg E_{\rm T}$. Since $\omega_n \gg E_{\rm T} = \hbar/2 \tau_{\rm dwell}$ for all $n$ in this high temperature regime, absorption is strong, $\tau_{\rm abs} \ll \tau_{\rm dwell}$. The formal correspondence between Matsubara frequency and absorption rate helps to understand that, to lowest order in $\tau_{\rm abs}/\tau_{\rm dwell} = E_{\rm T}/\omega$, the diagonal elements of $S(i\omega)$ are given by the reflection amplitudes of the tunnel barriers in the contacts, $S_{jj} = (1 - \Gamma_j)^{1/2}$, while the off-diagonal elements satisfy
\begin{equation} \label{eq:SP}
  \langle |S_{ij}(i \omega)|^2 \rangle = {\Gamma_i \Gamma_j \over \sum_{k=1}^{N} \Gamma_k} \int_0^{\infty} d\tau\, P_{ij}(\tau) \exp(-2 \omega \tau/\hbar), \ i \neq j.
\end{equation}
The function $P_{ij}$ is the classical distribution of dwell times for particles that enter the quantum dot through mode $j$ and exit through mode $i$. Because of the smallness of $\langle |S_{ij}(i \omega)|^2 \rangle = {\cal O}(E_{\rm T}/\omega)$, it is sufficient to keep only the lowest order term in an expansion of $\langle F(\omega) \rangle$ in the off-diagonal scattering matrix elements,
\begin{eqnarray}
  \langle F(\omega) \rangle = - {4e^2\over h} \sum_{i=1}^{N_1} \sum_{j=N_1+1}^{N} {2\alpha(i \omega)^2 \langle |S(i \omega)_{ij}|^2 \rangle \sin \phi \over [1 - \alpha(i \omega)^2(1-\Gamma_i)][1 - \alpha(i \omega)^2(1-\Gamma_{j})]}. \label{eq:Fexpand}
\end{eqnarray}
Eqs.\ (\ref{eq:IFsum})--(\ref{eq:Fexpand}) permit a semiclassical calculation of the average supercurrent in the non-ergodic regime for temperatures $k T \gg E_{\rm T}$, where random-matrix theory fails. The only input required is the classical distribution of dwell times.

On time scales greater than $\tau_{\rm erg}$, the distribution  $P_{ij}$ is exponential with the same mean dwell time $\tau_{\rm dwell} = \hbar/2 E_{\rm T}$ for all $i$, $j$:
\begin{equation}
   P_{ij}(\tau) = {2 E_{\rm T} \over \hbar} \exp(-2 E_{\rm T} \tau/\hbar). \label{eq:PTdwell}
\end{equation}
The non-chaotic dynamics on time scales shorter than $\tau_{\rm erg}$ enters through a non-universal form of $P_{ij}$ for $\tau \lesssim \tau_{\rm erg}$. We consider the case of a ballistic dynamics (size $L$ of the normal region much less than the mean free path $\ell$). The ergodic time $\tau_{\rm erg} \simeq L/v_{\rm F}$ is then a lower cutoff on $P_{ij}$, since the minimum dwell time $L/v_{\rm F}$ is the time needed to cross the system ballistically. A qualitative estimate of $\langle I \rangle$ is obtained if we set $P_{ij}(\tau) = 0$ for $\tau < L/v_{\rm F}$ and approximate it by Eq.\ (\ref{eq:PTdwell}) for larger times. Substitution of this dwell-time distribution into Eq.\ (\ref{eq:SP}) gives 
\begin{eqnarray}
  \langle |S_{ij}(i \omega)|^2 \rangle = {\Gamma_i \Gamma_j \over \sum_{k=1}^{N} \Gamma_k}  {E_{\rm T} \over \omega} \exp(-2 \omega L/\hbar v_{\rm F}), \ \omega \gg E_{\rm T},\ i \neq j.
\end{eqnarray}
We next compute $\langle F(\omega) \rangle$ from Eq.\ (\ref{eq:Fexpand}), replacing $\alpha(i \omega)$ by its value $-i$ for $\omega \ll \Delta$. The result is
\begin{subequations}\label{eq:Fmed}
\begin{eqnarray}
  \langle F(\omega) \rangle &=& {\tilde G E_{\rm T} \over \omega} \exp(-2 \omega L/\hbar v_{\rm F})\, {\sin \phi}, \\
  \tilde G &=& {2e^2 \over h} {\sum_{i=1}^{N_1} \sum_{j=N_1+1}^{N} 4\Gamma_i \Gamma_j (2-\Gamma_i)^{-1}(2-\Gamma_j)^{-1} \over \sum_{k=1}^{N} \Gamma_k}.
\end{eqnarray}
\end{subequations}%
Notice that $\tilde G = G$ for the case of high tunnel barriers ($\Gamma_j \ll 1$ for all $j$). We can now calculate the average supercurrent from Eq.\ (\ref{eq:IFsum}). Eq.\ (\ref{eq:Fmed}) is valid for $E_{\rm T} \ll \omega \ll \Delta$, $E_{\rm T} \ll \hbar v_{\rm F}/L \ll \Delta$, and is sufficient to determine the supercurrent in the temperature range $E_{\rm T} \ll k T \ll \Delta$. Substitution of Eq.\ (\ref{eq:Fmed}) into Eq.\ (\ref{eq:IFsum}) gives
\begin{subequations}  \label{eq:Iballistic}
\begin{eqnarray} \label{eq:Iballistica}
  \langle I \rangle &=& {\tilde G E_{\rm T} \over e}\,
  \lefteqn{ \sin \phi \, \ln \left( \hbar v_{\rm F}/ \pi k T L \right),}
  \hphantom{2 \sin \phi\, \left[ \ln \left(2 E_{\rm c}/E_{\rm T}\right) - c_{\rm Euler} \right],}
  \ \ E_{\rm T} \ll k T \ll \hbar v_{\rm F}/L \ll \Delta, \\
  \langle I \rangle &=& {\tilde G E_{\rm T} \over e}\,
  \lefteqn{2 \sin \phi\, \exp(-2 \pi k T L/\hbar v_{\rm F}),}
  \hphantom{2 \sin \phi\, \left[ \ln \left(2 E_{\rm c}/E_{\rm T}\right) - c_{\rm Euler} \right],}
  \ \ E_{\rm T} \ll \hbar v_{\rm F}/L \ll k T \ll \Delta.
\end{eqnarray}
\end{subequations}%
Eq.\ (\ref{eq:Iballistic}) has the same temperature dependence as the result of Ref.\ \onlinecite{AslamasovLarkinOvchinnikov} for the double-barrier SNS junction.

We now turn to low temperatures $kT \lesssim E_{\rm T}$. In this temperature regime, the Matsubara sum (\ref{eq:IFsum}) contains terms with $\omega_n \lesssim E_{\rm T}$, for which the off-diagonal scattering matrix elements $S_{ij}(i \omega_n)$ are not small and the approximation (\ref{eq:Fexpand}) is no longer valid. However, since $E_{\rm T} \ll \hbar/\tau_{\rm erg}$, these Matsubara frequencies are well within the validity range of random-matrix theory. Therefore, we can use the results of Sec.\ \ref{sec:3} to compute $\langle F(\omega) \rangle$ for $\omega \lesssim E_{\rm T}$ and the semiclassical formula (\ref{eq:Fexpand}) for $\omega \gtrsim E_{\rm T}$. These two results match at $\omega \simeq E_{\rm T}$, because the validity range $\omega \ll \hbar/\tau_{\rm erg}$ of random-matrix theory and the validity range $\omega \gg E_{\rm T}$ of the semiclassical theory overlap (assuming $\tau_{\rm erg} \ll \tau_{\rm dwell} = \hbar/2 E_{\rm T}$). 

For the case of high tunnel barriers, random matrix theory gives [cf.\ Eq.\ (\ref{eq:Itunnel})]
\begin{eqnarray}
  \langle F(\omega) \rangle = {G E_{\rm T} \sin \phi \over \sqrt{|\Omega(\phi)|^2 + \omega^2}},\ \ \omega \ll \hbar v_{\rm F}/L \ll \Delta, \label{eq:FItunnelNE1}
\end{eqnarray}
while the semiclassical formula (\ref{eq:Fexpand}) gives
\begin{eqnarray}
  \langle F(\omega) \rangle = {G E_{\rm T} \sin \phi \over \omega} \exp(-2 \omega L/\hbar v_{\rm F}),\ \ E_{\rm T} \ll \omega \ll \Delta. \label{eq:FItunnelNE2}
\end{eqnarray}
[The function $\Omega(\phi)$ was defined in Eq.\ (\ref{eq:Ed}).] The two results (\ref{eq:FItunnelNE1}) and (\ref{eq:FItunnelNE2}) have a common range of validity $E_{\rm T} \ll \omega \ll \hbar v_{\rm F}/L$, within which they can be matched. The result is a formula valid for all $\omega \ll \Delta$, for a ballistic quantum dot with high tunnel barriers:
\begin{eqnarray}
  \langle F(\omega) \rangle = {G E_{\rm T} \sin \phi \over \sqrt{|\Omega(\phi)|^2 + \omega^2}} \exp(-2 \omega L/\hbar v_{\rm F}),\ \ \omega \ll \Delta. \label{eq:FItunnelNE3}
\end{eqnarray}
After substitution of Eq.\ (\ref{eq:FItunnelNE3}) into Eq.\ (\ref{eq:IFsum}) we obtain the average supercurrent in the low-temperature regime,
\begin{eqnarray}
  \langle I \rangle &=& {G E_{\rm T}\over e} \sin \phi \,
  {\left[ \ln \left({ \hbar v_{\rm F} \over L E_{\rm T}
    \sqrt{1 - \gamma^2 \sin^2(\phi/2)}}\right) - c_{\rm Euler} \right],}
  \ \ k T \ll E_{\rm T} \ll \hbar v_{\rm F}/L \ll \Delta. \label{eq:Iballistic2}
\end{eqnarray}
[The parameter $\gamma$ was defined in Eq.\ (\ref{eq:y}).] The results (\ref{eq:Iballistic}) and (\ref{eq:Iballistic2}) cover the entire temperature range below $T_{\rm c}$.

\section{Conclusion} \label{sec:6}

In Table \ref{tab:1} we summarize the parametric dependence of the critical current at zero temperature on the three time scales $\tau_{\rm dwell}$, $\tau_{\rm erg}$, and $\hbar/\Delta$. We show the three new regimes for a weakly coupled normal region ($\tau_{\rm dwell} \gg \tau_{\rm erg}$), and have included for comparison also the two old regimes for a strongly coupled normal region ($\tau_{\rm dwell} \simeq \tau_{\rm erg}$). Apart from a logarithmic factor, the critical current is given by $I_{\rm c} \simeq (G/e) \min(\hbar/\tau_{\rm dwell},\Delta)$ in each of the five regimes. There is an additional logarithmic dependence on $\min(\tau_{\rm dwell}/\tau_{\rm erg},\tau_{\rm dwell} \Delta/\hbar)$ in two of the three new regimes. Upon raising the temperature, the critical current is suppressed at a characteristic temperature given by $\min(\hbar/\tau_{\rm erg},\Delta)$. At lower temperatures, $I_{\rm c}$ has a logarithmic $T$-dependence as long as $T \gtrsim \hbar/\tau_{\rm dwell}$ and becomes $T$-independent at still lower $T$.

\begin{table}
  \begin{tabular}{l|c|c|c}
  $I_c \times e/G$ & \multicolumn{2}{c|}{weak coupling ($\tau_{\rm dwell} \gg \tau_{\rm erg}$)} & strong coupling  \\
  ~ & ergodic ($\tau_{\rm erg} \ll \hbar/\Delta$) & non-ergodic ($\tau_{\rm erg} \gg \hbar/\Delta$) & ($\tau_{\rm dwell} \simeq \tau_{\rm erg}$) \\ \hline & & & \\
  short dwell time & $\Delta$ & --- & $\Delta$ \\ 
  ($\tau_{\rm dwell} \ll \hbar/\Delta$) & & & \\ \hline & & & \\
  long dwell time & $E_{\rm T} \ln\left(\mbox{\large ${\Delta \over E_{\rm T}}$}\right)$ & $E_{\rm T} \ln\left( \mbox{\large ${\hbar \over E_{\rm T} \tau_{\rm erg}}$}\right)$ & $E_{\rm T}$ \\
  ($\tau_{\rm dwell} \gg \hbar/\Delta$) & & &
  \end{tabular}
~\\

\caption{\label{tab:1} Parametric dependence of the zero-temperature critical current $I_{\rm c}$ on the three time scales $\tau_{\rm dwell}$, $\tau_{\rm erg}$, and $\hbar/\Delta$. The Thouless energy $E_{\rm T} = \hbar/2\tau_{\rm dwell}$.}
\end{table}

In this work, we did not address the sample-to-sample fluctuations of the supercurrent, but calculated only the ensemble average. For strongly coupled diffusive Josephson junctions ($\tau_{\rm dwell} \simeq \tau_{\rm erg}$, $L \gg \ell$), the root-mean-squared of the fluctuations is a factor $e^2/h G$ smaller than the average critical current \cite{Beenakker91,AltshulerSpivak}. Preliminary calculations in the ergodic regime indicate that the same is true for weakly coupled Josephson junctions ($\tau_{\rm dwell} \gg \tau_{\rm erg}$), i.e.\ the r.m.s.\ fluctuations of $I_{\rm c}$ are given by the entries in Table \ref{tab:1}, times $e/h$. 

We close with a remark on quantum dots with an {\em integrable} classical dynamics, such as rectangular or circular ballistic cavities. For energies $\varepsilon \lesssim E_{\rm T}$, the excitation spectrum of an integrable Josephson junction is quite different from its chaotic counterpart \cite{Melsen96}: The density of states $\rho(\varepsilon)$ of a chaotic cavity in contact with a superconductor shows a gap of size $E_{\rm T}$ around the Fermi level $\varepsilon=0$, while $\rho(\varepsilon)$ vanishes linearly when $\varepsilon \to 0$ for a rectangular or circular cavity. It is an interesting open problem, to compute the supercurrent through an integrable cavity and compare with the results for the chaotic case obtained in this paper.

\acknowledgments

We thank A.\ I.\ Larkin and Yu.\ N.\ Ovchinnikov for alerting us to the logarithmic temperature dependence of the supercurrent found in Ref.\ \onlinecite{AslamasovLarkinOvchinnikov}. This work was supported by the ``Stich\-ting voor Fun\-da\-men\-teel On\-der\-zoek der Ma\-te\-rie'' (FOM) and by the ``Ne\-der\-land\-se or\-ga\-ni\-sa\-tie voor We\-ten\-schap\-pe\-lijk On\-der\-zoek'' (NWO).


\begin{thebibliography}{99}
\bibitem{Marsh} A. M. Marsh, D. A. Williams, and H. Ahmed, Phys.\ Rev.\ B 
    {\bf 50}, 8118 (1994).
\bibitem{Takayanagi} H. Takayanagi, T. Akazaki, and J. Nitta, Phys.\ Rev.\
    B {\bf 51}, 1374 (1995).
\bibitem{Courtois1995} H. Courtois, Ph. Gandit, and B. Pannetier, 
    Phys.\ Rev.\ B {\bf 51}, 9360 (1995); {\bf 52}, 1162 (1995).
\bibitem{Courtois1996} H. Courtois, Ph. Gandit, D. Mailly, and B. Pannetier, 
    Phys.\ Rev.\ Lett.\ {\bf 76}, 130 (1996).
\bibitem{Charlat1996} P. Charlat, H. Courtois, Ph. Gandit, D. Mailly,
    A. F. Volkov, and B. Pannetier, preprint (cond-mat/9609182).
\bibitem{Mur} L. C. Mur, C. J. P. M. Harmans,
    J. E. Mooij, J. F. Carlin, A. Rudra,
    and M. Ilegems, Phys.\ Rev.\ B {\bf 54}, 2327 (1996).
\bibitem{Volkov95} A. F. Volkov, P. H. C. Magn\'ee, B. J. van Wees, and T. M
    Klapwijk, Physica C {\bf 242}, 261 (1995).
\bibitem{Volkov96} A. F. Volkov and H. Takayanagi, 
    Phys.\ Rev.\ Lett.\ {\bf 76},
    4026 (1996); Phys.\ Rev.\ B (to be published).
\bibitem{Wilhelm}
    F. K. Wilhelm, A. D. Zaikin, and G. Sch\"on, preprint (cond-mat/9608098).
\bibitem{Kulik} I. O. Kulik, Zh. Eksp. Teor.\ Fiz.\ {\bf 57}, 1745 (1969)
    [Sov.\ Phys.\ JETP {\bf 30}, 944 (1970)].
\bibitem{AslamasovLarkinOvchinnikov} L. G. Aslamasov, A. I. Larkin, 
    and Yu. N. Ovchinnikov, Zh.\ Eksp.\ Teor.\ Fiz.\ {\bf 55}, 323 (1968)
    [Sov.\ Phys.\ JETP {\bf 28}, 171 (1969)].
\bibitem{VanWeesTakayanagi} B. J. van Wees and H. Takayanagi, 
    in {\em Mesoscopic Electron Transport}, edited by 
    L. P. Kouwenhoven, G. Sch\"on, and L. L. Sohn, 
    NATO ASI Series E (Kluwer, Dordrecht, to be published).
\bibitem{Efetov83} K. B. Efetov, Adv.\ Phys. {\bf 32}, 53 (1983).
\bibitem{AAAS} A. V. Andreev, O. Agam, B. D. Simons,
    and B. L. Altshuler, Phys.\ Rev.\ Lett.\ {\bf 76}, 3967 (1996).
\bibitem{AS} B. L. Altshuler and B. I. Shklovski\u{\i}, 
    Zh.\ Eksp.\ Teor.\ Fiz. {\bf 91},
    220 (1986) [Sov.\ Phys.\ JETP {\bf 64}, 127 (1986)].
\bibitem{Likharev} K. K. Likharev, Rev.\ Mod.\ Phys.\ {\bf 51}, 101 (1979).
\bibitem{Svidzinskii} A. V. Svidzinski\u{\i}, T. N. Antsygina, 
    and E. N. Bratus',
    Zh.\ Eksp.\ Teor.\ Fiz.\ {\bf 61}, 1612 (1971)
    [Sov.\ Phys.\ JETP {\bf 34}, 860 (1972)].
\bibitem{Melsen96} J. A. Melsen, P. W. Brouwer, K. M. Frahm, and C. W. J.
    Beenakker, Europhys.\ Lett.\ {\bf 35}, 7 (1996); Physica Scripta, 
    to be published (cond-mat/9607036).
    Phys.\ Rev.\ B {\bf 54}, 9443 (1996).
\bibitem{Beenakker91} C. W. J. Beenakker, Phys.\ Rev.\ Lett.\ 
    {\bf 67}, 3836 (1991);
    {\bf 68}, 1442 (E) (1992).
\bibitem{Beenakker92} 
    C. W. J. Beenakker, in {\em Transport Phenomena in Mesoscopic Systems}, 
    edited by H. Fukuyama and T. Ando (Springer, Berlin, 1992). 
\bibitem{Andreev} A. F. Andreev, Zh.\ Eksp.\ Teor.\ Fiz.\ {\bf 46}, 1823
    (1964) [Sov.\ Phys.\ JETP {\bf 19}, 1228 (1964)].
\bibitem{DoronSmilansky} E. Doron and U. Smilansky, Phys.\ Rev.\ Lett.\ 
    {\bf 68}, 1255 (1992).
\bibitem{BeenakkerVanHouten} C. W. J. Beenakker and H. van Houten, in 
    {\em Nanostructures and Mesoscopic Systems}, edited by W. P. Kirk 
    (Academic, New York, 1992).
\bibitem{MahauxWeidenmueller} C. Mahaux and H. A. Weidenm\"uller, 
   {\em Shell-Model Approach to Nuclear Reactions} 
   (North-Holland, Amsterdam, 1969).
\bibitem{VWZ} J. J. M. Verbaarschot, H. A. Weidenm\"uller, and M. R. Zirnbauer,
    Phys. Rep. {\bf 129}, 367 (1985).
\bibitem{Mehta} M. L. Mehta, {\em Random Matrices} (Academic, New York, 1991).
\bibitem{IWZ} S. Iida, H. A. Weidenm\"uller, and J. A. Zuk, Phys.\ Rev.\ 
    Lett.\ {\bf 64}, 583 (1990); Ann. Phys. (N.Y.) {\bf 200}, 219 (1990).
\bibitem{Brouwer95} P. W. Brouwer, Phys.\ Rev. B {\bf 51}, 16878 (1995).
\bibitem{Frahm96} K. M. Frahm, P. W. Brouwer, J. A. Melsen, and C. W. J. 
    Beenakker, Phys.\ Rev.\ Lett.\ {\bf 76}, 2981 (1996).
\bibitem{Pandey} A.\ Pandey, Ann.\ Phys.\ (N.Y.) {\bf 134}, 110 (1981).
\bibitem{BrezinZee} E.\ Br\'ezin and A.\ Zee, Phys.\ Rev.\
    E {\bf 49}, 2588 (1994).
\bibitem{Pastur} L. A. Pastur, Teor.\ Mat.\ Fiz.\ {\bf 10}, 102 (1972) 
    [Theor.\ Math.\ Phys. {\bf 10}, 67 (1972)].
\bibitem{Dwell} The dwell-time is defined as $\tau_{\rm dwell} = 
    (\hbar/i N) \partial\ln \det S(\varepsilon)/\partial \varepsilon$, 
    see E. P. Wigner, Phys.\ Rev.\ {\bf 98}, 145 (1955) and 
    F. T. Smith, Phys.\ Rev.\ {\bf 118}, 349 (1960).
\bibitem{BrouwerBeenakker1996a} P. W. Brouwer and C. W. J. Beenakker, 
    J.\ Math.\ Phys.\ {\bf 37}, 4904 (1996).
\bibitem{BarangerMello94} H. U. Baranger and P. A. Mello, 
    Phys.\ Rev.\ Lett.\ {\bf 73}, 142 (1994).
\bibitem{JPB} R. A. Jalabert, J.-L. Pichard, and C. W. J. Beenakker, 
    Europhys.\ Lett.\ {\bf 27}, 255 (1994).
\bibitem{AltshulerSpivak} B. L. Altshuler and B. Z. Spivak, 
   Zh.\ Eksp.\ Teor.\ Fiz.\ {\bf 92}, 609 (1987) 
   [Sov.\ Phys.\ JETP {\bf 65}, 343 (1987)].
\end{thebibliography}
\end{document}